\documentclass[preprint,aps,showpacs,amsmath,amssymb]{revtex4}
\usepackage{graphicx} 
\usepackage{dcolumn}
\usepackage{bm}
\begin{document}
\title{Inverse flux quantum periodicity of magnetoresistance oscillations
in two-dimensional short-period surface superlattices}
\author{X. F. Wang,$^\dagger$ P. Vasilopoulos$^{\diamond}$, and F. M. Peeters$^\star$
\ \\
\ \\}

\address{$^{\dagger, \diamond}$Concordia University Department of Physics\\
1455 de Maisonneuve  Ouest, Montr\'{e}al, Qu\'{e}bec, Canada, H3G 1M8\\
\ \\
$^{\star}$Departement Natuurkunde Universiteit Antwerpen
(UIA),\\
Universiteitsplein 1, B-2610, Belgium
\ \\}
\date{\today} 
\begin{abstract}
Transport properties of the two-dimensional electron gas (2DEG)
are considered in the presence of a perpendicular magnetic field $B$
and of a {\it weak} two-dimensional (2D) periodic potential
modulation in the 2DEG plane. The symmetry of the latter is
rectangular or hexagonal.
The well-known solution of the corresponding tight-binding equation shows
that each Landau level splits into several subbands when a rational number of flux quanta $h/e$ pierces the unit cell
and that the corresponding gaps
are exponentially small. Assuming the latter are closed due to
disorder gives analytical wave functions   and simplifies considerably the evaluation 
of the magnetoresistivity tensor $\rho_{\mu\nu}$. The relative phase of the oscillations in $\rho_{xx}$ and $\rho_{yy}$ depends on the modulation periods involved. For a 2D 
modulation with a {\bf short} period $\leq 100$ nm,  in addition to the Weiss oscillations the
collisional contribution to the conductivity and consequently the 
tensor $\rho_{\mu\nu}$ show {\it prominent peaks
when one  flux quantum $h/e$ passes through an integral number of  unit cells}
in good agreement with recent experiments.
For periods $300- 400$ nm long  used in early experiments, these peaks occur at
fields $10-25$ times smaller than those of the Weiss oscillations
and are not resolved.
\end{abstract}  


\maketitle

\section{ INTRODUCTION}
In the last decade the magnetotransport of the 2DEG, 
subjected to periodic potential
modulations, has attracted considerable experimental \cite{1}
and theoretical \cite{2,3} attention. For one-dimensional (1D) modulations novel
oscillations of the magnetoresistivity tensor $\rho_{\mu\nu}$
have been observed, at low magnetic fields $B$, distinctly different
in period and temperature dependence from the usual Shubnikov-de Haas (SdH)
ones observed at higher $B$. These novel oscillations reflect the commensurability
between two length scales: the cyclotron diameter at the Fermi level
$2R_c = 2\sqrt{2\pi n_e} \ell^2$, where $n_e$ is the electron
density, $\ell$ the magnetic length, and $a$ the period of the potential
modulation. The situation is similar but less  clearcut for 2D modulations
from both a theoretical \cite{4}-\cite{6} and an experimental \cite{6,8}  point of view.
To date most of the experimental results pertinent to 2D modulations 
\cite{6}-\cite{8} 
with square or hexagonal symmetry have indicated strongly that the 
predicted \cite{9}
fine structure of the Landau levels is not resolved. Magnetotransport theories
pertinent to this case are rather limited \cite{4},\cite{6,7} in contrast with those for 
1D modulations. 

Recent observations \cite{8} call for 
additional theoretical work since they could not be fully explained by earlier
semiclassical theories \cite{4}. In this
paper we develop, along the lines of Ref. 3, the relevant quantum mechanical
magnetotransport theory of the 2DEG for precisely the case that
the fine structure of the Landau levels is not resolved.
Our goal is  to explain recent experimental results \cite{13} on 2D, short-period 
($a\sim 1000$\AA) surface superlattices 
with mobility $\mu$   in the intermediate range,
i.e., $\mu \sim 100$ m$^2$/Vs. 
The symmetry of the 2D modulation is taken to be rectangular or hexagonal. 
A brief semiclassical account, pertinent to the former symmetry, was reported in Ref. 4 b).
New magnetoresistance oscillations are found to occur {\it when one flux quantum $h/e$ passes through an integral number of unit cells} as was recently observed experimentally \cite{13}.
Here we show that these oscillations   result 
from the interplay between band conduction and
collisional conduction. 
A new contribution to the latter
opens up 
as   hopping between cyclotron orbits which are separated
by an integral multiple  of the modulation period 
and have the
same position relative to the modulation lattice. This contribution 
is appreciable only in short-period superlattices and accordingly could not be resolved
in early experiments on long-period  superlattices.

In the next section we derive the one-electron eigenfunctions and
eigenvalues for rectangular and hexagonal modulations; we also present
the density of states. The analytical and numerical results for the
corresponding conductivity or resistivity components are presented
in Sec. III. Numerical results are given in Sec. IV and 
concluding remarks in Sec. V.

\section{ Eigenvectors, eigenvalues, and density of states}
 
We consider a 2DEG, in the $(x,y)$ plane, in the presence of
a perpendicular magnetic field ${\bf B} = B\hat{z}$ and of a 2D
periodic potential modulation $U(x,y)$. The electrons are
considered as free particles with an effective mass $m^*$.
In the absence of the modulation the normalized one-electron
eigenfunction, in the gauge ${\bf A} = (0,Bx,0)$, is given
by $e^{ik_yy}\phi_n(x + x_0)/\sqrt{L_y}$ where $\phi_n(x + x_0)$
is the well-known harmonic oscillator function, centered at
$-x_0= -\ell^2k_y$, and $L_y$ is the sample's width.
 
In the presence of a sinusoidal 1D modulation one can use
perturbation theory \cite{2,3} to evaluate the energy spectrum and
eigenfunctions. Alternatively, one can use a tight-binding scheme,
along the lines of Ref. \cite{9}, and look for solutions of the one-electron
Hamiltonian $H^0 = ({\bf p} + e{\bf A})^2/2m^* + V_x\cos(K_xx)$ that
are linear combinations of the unperturbed $(V_x = 0)$ ones:
$\mid \varphi_{nk_y} > =  \sum_p A_p\mid n,k_y + pG>$, where $G$ is a
suitable wave vector introduced here for convenience and $|n, k_y+pG>$ is the
unperturbed state ($ K_x = 2\pi/a_x$, and $a_x$ is
the modulation period along $x$). 
As in Ref. \cite{9} we take $G\equiv K_y = 2\pi/a_y$ with $a_y$  
the modulation period along $y$. The summation over $p$ has to be
extended to all integer $p$ values such that $-L_x/2\ell^2
\leq k_y + pK_y \leq L_x/2\ell^2$, where $L_x$ is the length.
For $p=0$ we have the limits for $k_y$ as $-a_x/2\ell^2 \leq k_y \leq a_x/2\ell^2$.
Then the tight-binding equation $<n,k_y + pK_y\mid H^0 - E|\varphi_{nk_y}> = 0$, in
which mixing of different Landau levels $n$ is neglected, gives acceptable
solutions for the coefficients $A_p$ as $A_p = A_0\exp(i\xi p)$.
The new states are labeled with the additional quantum number $\xi$
$(0 \leq \xi \leq 2\pi )$: $\mid \psi_{nk_y\xi} > = A_0 \sum_p \exp(i\xi p)
|n,k_y + pK_y>$. The orthonormality condition $<\varphi_{nk_y\xi}\mid
\varphi_{nk_y\xi'}> = \delta_{\xi\xi '}$ gives
$\xi = 2\pi\nu\ell^2 K_y/L_x \rightarrow k_x\ell^2 K_y$, $\nu$ being
an integer, and $A_0 = \ell(K_y/L_x)^{1/2}$ by normalization.
The energy spectrum obtained in this way is the same as that
obtained by perturbation theory \cite{2,3}.

We will now use this information to obtain the corresponding
eigenvectors and eigenvalues for a 2D modulation potential. 

\subsection{ Rectangular symmetry}

We assume the following one-electron Hamiltonian
\begin{equation}
H^0 = ({\bf p} + e{\bf A})/2m^* + V_x\cos(K_x x) + V_y\cos(K_y y)
\end{equation}
where $K_\mu = 2\pi/a_\mu (\mu = x,y)$; 
$a_x$ and $a_y$ are the
periods along the $x$ and $y$ directions, respectively.

In the gauge chosen, ${\bf A} = (0,Bx,0)$, the second term of Eq. (1)
is not diagonal in $k_y$ and therefore $\mid n,k_y>$ is not a
convenient basis set. But as in the 1D case we can look for solutions
of Eq. (1) in the form $\mid \varphi_{n,k_y} > = \sum_p A_p \mid n,k_y + pK_y>$
as described above. This choice of eigenfunctions 
is also suggested by the fact that $V_y\cos K_yy$
connects the unperturbed state $\mid nk_y>$ with only the states
$\mid n,k_y \pm K_y>$.
In this case the equation $<n,k_y + pK_y\mid H^0 - E\mid \varphi_{n,k_y}> = 0$,
in which mixing of different Landau levels is neglected,  takes the form
\begin{equation}
V_x F_n(u_x )\cos(2\pi p \alpha + K_x x_0) A_p + {{}_1\over {}^2}
V_y F_n(u_y)(A_{p+1} + A_{p-1}) = (E - E_n)A_p,  
\label{harper}
\end{equation}
where $\alpha = 2\pi\ell^2/a_xa_y$, $E_n = (n + 1 /2)
\hbar \omega_c$ is the ``unperturbed'' eigenvalue and 
$\omega_c = | e| B/ m^*$
the cyclotron frequency. Further, $F_n(u_\mu) = \exp(-u_\mu/2)
L_n(u_\mu), \ L_n(u_\mu)$ is the Laguerre polynomial,
and $u_\mu = \ell^2K^2_\mu/2$.

\begin{figure}[tpb]
\vspace{-1.5cm}
\includegraphics*[width=80mm,height=70mm]{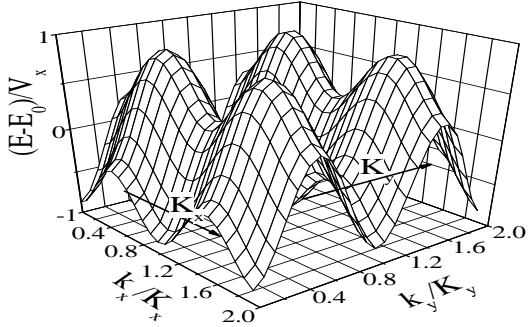}
\vspace{-1.5cm}
\caption{Energy spectrum  as a function of the  wave vectors $k_x$ and $k_y$ for
$n=0,\ \alpha=1$, $V_x=2V_y=1$ meV.  The  modulation wave vectors $K_x$ and $K_y$ are shown
by the thick arrows.} 
\label{fig1}
\end{figure}

The solution of Eq. (2) gives the eigenvalues $E$ and the eigenvectors
$A_p$. We see immediately that for $\alpha$  integer 
the equation admits the exponential solutions $A_p = A_0 e^{i\xi p}$,
with $A_0$ and $\xi$  given above. This is also the case for those
values of $\alpha$ for which $F_n(u_x)$ vanishes since
$u_x = 2\pi^2\ell^2/a^2_x = \pi(a_y/a_x)\alpha$. In the
former case we have
\begin{equation}
E_{nk\xi} = E_n + V_x F_n (u_x) \cos(K_xx_0) +
V_y F_n(u_y) \cos \xi  
\end{equation}
and in the latter
\begin{equation}
E_{nk\xi} = E_n + V_y F_n(u_y)\cos\xi, 
\end{equation}
where $\xi = (2\pi\nu/L_x)\ell^2K_y \equiv \ell^2K_yk_x$. 
In both cases the unperturbed Landau levels broaden into bands (with a
bandwidth equal to $2(V_x |F_n(u_x)| + V_y |F_n(u_y)|)$ and
$2V_y |F_n(u_y)|$, respectively), that oscillates with magnetic
field $B$ and (large) index $n$, cf. Refs. 2 and 3.  The energy spectrum given by Eq. (3),
plotted in Fig. \ref{fig1} for $n=0,\ \alpha=1$, $V_x=2V_y=1$ meV, is a periodic function of
$k_x$ and of $k_y$ since $\xi = \ell^2K_yk_x$ and $x_0= \ell^2k_y$.
Notice that the arguments
of the cosines in Eq. (3) can be shifted by $2\pi\alpha$, $\alpha$ integer. 

One important consequence of this nonzero  bandwidth is that the mean
velocities $v_x$ and $v_y$, which vanish 
in the absence of
modulation, are now finite: from
$v_\mu = d E_{nk\xi}/\hbar dk_\mu, \mu = x,y$, we obtain
\begin{equation}
v_x = - (\ell^2K_yV_y/\hbar)F_n(u_y) \sin\xi,  
\end{equation}
and
\begin{equation}
v_y = - (\ell^2 K_xV_x/\hbar)F_n(u_x)\sin(K_xx_0).
\end{equation}
Eqs. (5) and (6) lead to a finite diffusion or band
conductivity which is absent when the modulation is not present, cf.
Sec. II.

Equation (2) is the same as Harper's equation but the coefficients
$V_\mu F_n(u_\mu)$ depend on the magnetic field.
For $B$ values other than those pertaining to Eq. (3)
it has been shown by Hofstadter \cite{10}a for the case of constant coefficients
and by Claro and Wannier \cite{10}b,
for the case that the latter depend on $B$ (hexagonal
modulation), that the energy spectrum resulting from the numerical
solution of Eq. (2), i.e., $E$, shows, when $E$ is measured in units
of $V_\mu F_n(u_\mu )$, 
a nontrivial structure:
for $\alpha = i/j$, $i,j$ being integers, each Landau level is
split into $j$ subbands and Eq.  (2) is periodic with period $j$. 
Here, in view of the reported experiments \cite{6}-\cite{8}
which did not indicate that this fine structure of the energy
spectrum was resolved, we will assume that this is indeed the case,
i.e., that in samples of not exceptionally high mobilities, such as those of Ref. 5, the
small gaps mentioned above are closed due to disorder and justify the assumption
below, see subsection C.
That is, we assume that the Landau levels are bands. Now from the
numerical solution of Eq. (2) we know that the bandwidth corresponding
to $\alpha = i/j$ cannot exceed the value obtained
from Eq. (3) or Eq. (4). Therefore, for computational convenience,
we will assume that  the energy spectrum
is given by Eq. (3) and the eigenfunctions by
$\mid \psi_{nk_y\xi}> = A_0 \sum_p e^{ip\xi}\mid n,k_y + pK_y>$.

\begin{figure}[tpb]
\vspace{-1.5cm}
\includegraphics*[width=130mm,height=130mm]{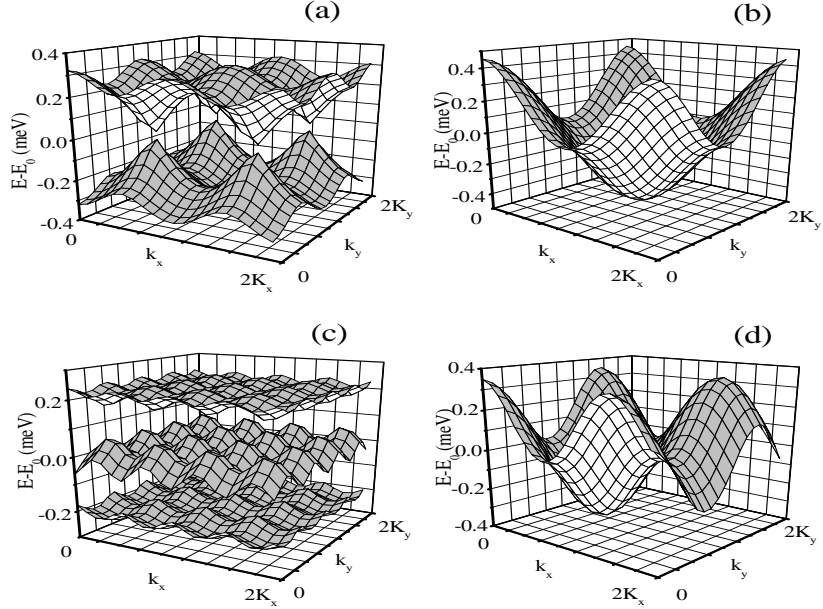}
\vspace{-1.5cm}
\caption{Energy spectrum, obtained from the exact solution of 
Eq. (\ref{harper}), in (a) and (c), as a function of the  wave vectors $k_x$ and $k_y$
for  $n=0$. The corresponding approximate spectrum, given by Eq. (3), is shown in (b) and (d).
In (a) and (b) we have $\alpha=1/2$ and $\alpha=2/3$, in (c) and (d) $\alpha=2/3$.
The periods $a_x=a_y=800$\AA\ pertain to the experiment of Ref. \cite{13}.}
\label{fig14}
\end{figure}

In Fig. \ref{fig14} we compare the energy spectrum obtained by exactly solving
Eq. (\ref{harper}) with the  one given by Eq. (3). We do so  because in the conductivity
calculations we will use  Eq. (3), as an approximation that will be justified, for all magnetic fields or values  of $\alpha$. The exact spectrum for $\alpha=i/j$
is composed of j minibands. It's dependence on $k_x$, which does not appear in Eq. (2), is obtained by introducing appropriate new basis states, in the manner of Ref. \cite{10a}, with
$|k_x|\leq \pi/ia_x$ and  $|k_y|\leq \pi/a_y$ restricted in the magnetic Brillouin zone. 
As can be seen, the two spectra are quite different from each other. The corresponding difference  in the density of states is much weaker if a small
broadening is included, see subsection C  below. For  
$\alpha$ integer, however, the exact spectrum and that given by Eq. (3) coincide; the result is shown in Fig. 1. 

There are two alternative, {\it approximate} ways to obtain
Eq. (3). First, we take $V_y \approx 0$ and use the corresponding
1D tight-binding states $\mid n, k_y, \xi>$ to obtain the energy
spectrum given by $E_n + V_xF_n (u_x)\cos(K_xx_0)$.
We then use first-order perturbation theory, involving the
states $\mid nk_y\xi>$, to evaluate the energy correction to this
spectrum due to the term $V_y\cos K_yy$ for $V_y\ll \hbar \omega_c +
V_x$; the result is identical with that given by Eq. (3).
Secondly, since these new oscillations of the magnetoresistance
have been observed in weak magnetic fields and for weak modulations,
we attempt a classical evaluation of the correction to the 
unperturbed energy $E_n$ by the modulation
$V_x\cos K_x x + V_y\cos K_y y$ using the classical equations
of motion $x(t) = x_0 + R_c \sin(\omega_c t + \varphi), y(t) = y_0 + R_c
\cos(\omega_ct + \varphi)$; $x_0$ and $y_0$ are the classical center
coordinates, $R_c$ is the cyclotron radius, $\omega_c = |e|
B/m^*$, and $\varphi$ is a phase factor. Without loss of generality
we may take $\varphi = 0$. Then, if $T$ is the period of the
cyclotron motion, a straightforward evaluation gives
\begin{eqnarray}
\nonumber
<U> &=&(1/T) \int^{T/2}_{-T/2} dt  [V_x\cos K_xx(t) +
V_y\cos K_yy(t)]\\* 
&&= V_x J_0(K_xR_c)\cos K_x x_0 + V_yJ_0(K_y R_c) \cos K_y y_0,  
\end{eqnarray}
where $J_0(x)$ is the Bessel function of order zero. In the
weak magnetic field limit $K_\mu R_c \gg 1$, Eq. (7) reduces
to Eq. (3) for large $n$, i.e.,  for weak $B$. It is obvious that
these three {\it approximate} ways of deriving the energy spectrum
do not ``see'' its fine structure resulting from an exact numerical
evaluation of the finite-difference Eq. (2) for $\alpha = i/j$.
Therefore, they are applicable  if the corresponding small gaps are closed
due to disorder. 

\subsection{ Hexagonal symmetry}

We assume that the Hamiltonian is given by

\begin{equation}
H^0 = ({\bf p} + e{\bf A})^2/2m^* + V_x \cos K_xx \cos K_yy +
V_y 
(1+ \cos 2K_yy)/2.  
\end{equation}
For $V_x = V_y = V_0$ this reduces to the model studied experimentally
by Fang and Stiles \cite{6}. If $x$ and $y$ are interchanged 
the energy spectrum, with $K_y = 2\pi/a$ and $K_x = 2\pi/\sqrt{3} a$,
of the corresponding tight-binding equation 
has been studied numerically, for all values of the
magnetic field, by Claro
and Wannier \cite{10} and has the same structure as that of the square symmetry.
Here, in line with the case of rectangular symmetry, we assume that
the small gaps of the energy spectrum are closed due to disorder and
 use again the tight-binding  description of Sec. II  {\bf A}. 
 Corresponding to Eq. (2) we now obtain 
\begin{eqnarray}
\nonumber 
 {1\over 2}V_xF_n (u_x + u_y)[\cos(2\pi p \alpha +\gamma)A_{p+1} &+& 
 \cos(2\pi p \alpha - \gamma)A_{p-1}]
 +{V_y\over 4}  F_n (4u_y)[A_{p+2} + A_{p-2}]\\*
&&= (E - E_n - {{}_1\over{}^2}V_y)A_p,  
\end{eqnarray}
where $\gamma = K_x \ell^2(k_y + K_y/2)$. When $\alpha$ is  integer  
  Eq. (9) has the solution $A_p = A_0 \exp(i\xi p)$ with
$A_0$ and $\xi$ given in A and the eigenvalue  $E$ is given by
\begin{equation}
E_{nk\xi} = E_n +{1 \over  2}V_y + V_xF_n (u_x + u_y)\cos\gamma\cos\xi + 
{1\over 2}V_yF_n(4u_y)\cos 2\xi, 
\end{equation}
where $\xi = \ell^2 K_y k_x$. We notice that for $K_x = 2\pi/a$ and
$K_y = 2\pi\sqrt{3}a$, i.e. the usual hexagonal modulation, we
have $F_n(u_x + u_y) = F_n (4u_y) = F_n (8 \pi^2\ell^2/3a^2)$.

As in the rectangular case, we see that the Landau levels have broadened
into bands with a bandwidth equal to $(2V_x |F_n(u_x + u_y) +
V_y|F_n(4u_y)|)$ that oscillates with magnetic field and
(large) $n$. Again the mean velocities $v_x$ and $v_y$ are finite
\begin{equation}
v_x = -(V_y\ell^2 K_y/\hbar)F_n(4u_y)\sin 2\xi
- (V_x\ell^2 K_y/\hbar) F_n(u_x + u_y) \cos\gamma\cos\xi, 
\end{equation}
\begin{equation}
v_y = - (V_x \ell^2 K_x/\hbar) F_n(u_x + u_y) \sin\gamma\sin\xi; 
\end{equation}
this has important consequences for transport and will be detailed in the
next section. 

\subsection{ The density of states}

The energy spectra given by Eqs. (3) and (10) are qualitatively
different  from  the unmodulated spectrum, given by $E_n$, and
from the corresponding 1D modulation spectrum given by
$E_n + F_n(u_x)\cos K_x x_0$. These differences are also
reflected in the density of states (DOS) defined by $D(E) = 2
\sum_{nk_y\xi} \delta(E - E_{nk_y\xi})$.
 For a 2D modulation with rectangular symmetry,
    corresponding to Eq. (3), the DOS becomes 
\begin{equation}
D(E ) = D_0 \sum_{n=0}^{\infty} \int^{2\pi}_0 d\xi 
\{ \left[ V_xF_n(u_x)\right]^2
-\left[E - E_n - V_y F_n(u_y)\cos\xi\right]^2 \}^{-1/2},
 \end{equation}

\noindent while for the one with hexagonal symmetry, corresponding
    to Eq. (10), the DOS is given by 
\begin{equation}
D(E) = D_0\sum_{n=0}^{\infty} \int^{2\pi}_0 d\xi 
\left\{ \left[ V_xF_n(u_x + u_y)\cos\xi\right]^2 - \left(E - E_n - 
{V_y \over 2} [1+F_n (4u_y)\cos 2\xi ]\right)^2 \right\}^{-1/2}, 
\end{equation}
where $D_0 = L_y L_x/\pi^3\ell^2$.
The quantities within the curly brackets  in Eqs. (13) and (14)   must be positive.

\begin{figure}[tpb]
\vspace{-1.5cm}
\includegraphics*[width=120mm,height=120mm]{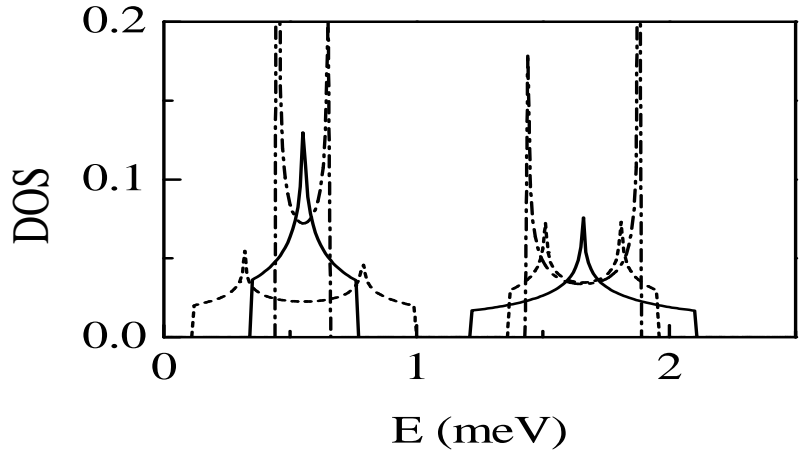}
\vspace{-1.5cm}
\caption{Density of states versus energy for $V_x=V_y=0.5$ meV, $a_x=800$\ \AA\ 
with   $a_y=800$\ \AA\  and $a_y=1600$\ \AA\ for the solid and dotted curves, respectively.
The dash-dotted curve is  the result for a 1D modulation  along the x-direction with the
same period and modulation strength as  in the 2D case.
The  magnetic field is $B=0.64$ T.}
\label{fig2}
\end{figure}

In Fig. \ref{fig2} we plot the DOS, given by Eq. (13), for various
values of the parameters
$a_x, a_y, V_x$, and $V_y$. For comparison we also show the 
DOS (dash-dotted curve) corresponding to the 1D
modulation. The latter exhibits van Hove singularities at the
edges of each Landau level (band) reflecting the 1D nature of the
electron motion in this band, since $v_x \neq 0$ while
$v_y = 0$, cf. Ref. 2 and 3.
This is not the case for the 2D modulation: the electron motion
is two-dimensional, since both $v_x$ and $v_y$ are different from zero, cf.
Eqs. (5) and (6). That is, in the 2D case the DOS is finite, see also Ref. 4 b).
As shown there, the DOS is qualitatively the same as the one shown in Fig. \ref{fig2}
if the periods
are the same and the strengths are varied. This can be immediately deduced from the factors
$V_xF_n(u_x)$ and $V_y F_n(u_y)$ that appear in Eqs. (3) and (11).
The DOS for the hexagonal modulation is not shown since it's similar to the one
shown in Fig. \ref{fig2}.

\begin{figure}[tpb]
\vspace{-1.5cm}
\includegraphics*[width=80mm,height=90mm]{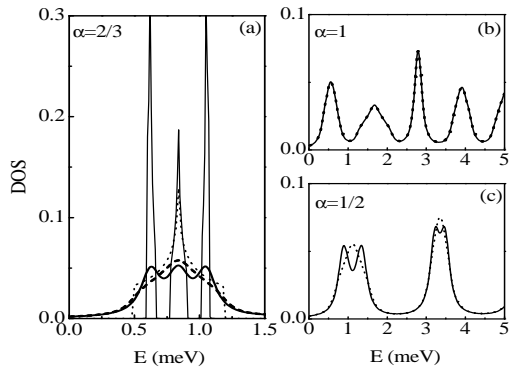}
\vspace{-1.5cm}
\caption{(a) Density of states versus energy for $\alpha=2/3$ and   
energy level width 
$\Gamma=0$ (thin curves) and $\Gamma=1K$ (thick curves). The solid and   dotted curves are, respectively, the exact and  approximate 
results. The DOS for $\alpha=1$ in (b) and $\alpha=1/2$ in (c) is plotted, respectively, 
for $\Gamma=1.1$K  and $\Gamma=1.5$K 
obtained from $\Gamma=(e\hbar/m^*)\sqrt{B/\pi\mu}$. 
The parameters used are the same as those in Fig. \ref{fig14}.}
\label{fig15}
\end{figure}

In Fig. \ref{fig15} we compare the DOS obtained from the exact energy spectrum with that   obtained using Eq. (3). In this comparison we include  a level
broadening by replacing $\delta(E)$ in the definition of $D(E)$ by
$\pi\Gamma/(E^2+\Gamma^2)$. In Fig. \ref{fig15}(a) we show the influence of the level
broadening on the DOS for $\alpha=2/3$ and different values of $\Gamma$ specified in the caption. As can be seen, 
the subband structure  disappears with increasing $\Gamma$ and the exact and approximate
result   approach each other. That is,  the gaps between the minibands in each
Landau level are closed  with increasing level broadening. Notice that this happens for
quite small values of $\Gamma$ compared to the cyclotron energy which is about
1 meV in this example. As shown in (c), the same behavior of the DOS occurs for $\alpha=1/2$. Notice also that this closeness between the exact and approximate DOS occurs despite the drastic difference in the corresponding energy spectra shown in Fig. 2.
In addition, as shown in (b), for integer $\alpha$ 
the exact and  approximate results for the DOS are identical.

\section{ Transport coefficients}
\subsection{  Basic expressions}

For weak electric fields $E$, i.e. for linear responses, and weak
scattering potentials the conductivity tensor $\sigma_{\mu\nu}(\omega )$
in the one-electron approximation, has been evaluated in detail in Ref. 3: 
$\sigma_{\mu\nu}(\omega) = \sigma^d_{\mu\nu}(\omega) +
\sigma^{nd}_{\mu\nu}(\omega),\mu,\nu = x,y$.  The contribution $\sigma^d_{\mu\nu}(\omega)$
stems from the diagonal part of the density operator $\rho$.
In a suitable basis $<J^d_\mu> = Tr(\rho^dJ_\mu) = \sigma^d_{\mu\nu}E_\nu$,
where $J_\mu$ is the current density, and $\sigma^{nd}_{\mu\nu}(\omega)$
comes from the nondiagonal part of  $\rho(\rho = \rho^d + \rho^{nd})$.
In general $\sigma^d_{\mu\nu}(\omega) = \sigma^{dif}_{\mu\nu}(\omega) +
\sigma^{col}_{\mu\nu}(\omega)$, where $\sigma^{dif}_{\mu\nu}(\omega)$
indicates diffusive contributions and $\sigma^{col}_{\mu\nu}(\omega)$ 
collisional contributions. For the diffusive contribution we have 
\begin{equation}
\sigma^{dif}_{\mu\nu} (0) = {\beta e^2 \over \Omega} \sum_\zeta
f_\zeta (1 - f_\zeta)\tau (E_\zeta)v^\zeta_\mu v^\zeta_\nu,
\end{equation}
provided that the scattering is elastic or quasielastic, and for the
collisional one

\begin{equation}
\sigma^{col}_{\mu\nu} (0) = {e^2 \over 2\Omega} \sum_{\zeta\zeta'}
f_\zeta(1 - f_{\zeta'}) W_{\zeta \zeta'}
 (\alpha^\zeta_\mu - \alpha^{\zeta'}_\mu)^2,
\end{equation}
for both  elastic $(f_\zeta = f_{\zeta'})$ and inelastic
$(f_\zeta \neq f_{\zeta'})$ scattering. $W_{\zeta \zeta'}$ 
is the transition rate between the
unperturbed one-electron states $\mid \zeta>$ and $\mid \zeta'>$,
$\Omega$   the volume of the system, $e$ the electron charge,
$\tau(E_\zeta)$ the relaxation time, and $\alpha^\zeta_\mu = <\zeta
\mid r_{\mu} \mid \zeta>$ the mean value of the $\mu$-component of the
position operator when the electron is in state $\mid\zeta>$ and
has velocity $v^\zeta_\mu = < \zeta\mid v_\mu\mid\zeta>$. Equation (15)
describes transport through extended states whereas Eq. (16) deals with 
transport through localized states and is absent in semiclassical
treatments.

The nondiagonal contribution $\sigma^{nd}_{\mu\nu}(\omega)$ to the conductivity
is given by
\begin{eqnarray}
\nonumber
\sigma^{nd}_{\mu\nu}(\omega) = {2i\hbar e^2 \over \Omega}
&&\sum_{\zeta\neq \zeta'} f_\zeta(1 - f_{\zeta'})  < \zeta\mid v_\mu\mid \zeta' > 
<\zeta'\mid v_\nu\mid\zeta >\\*
&&\times {1 - e^{\beta(E_\zeta-E_{\zeta'})} \over 
E_\zeta-E_{\zeta'}} \lim_{\epsilon \rightarrow 0}
{1 \over E_\zeta - E_{\zeta'} + \hbar\omega + i\epsilon }.  
\end{eqnarray}
If we use the identity $f_\zeta(1 - f_{\zeta'}) \exp\left[ \beta
(E_\zeta - E_{\zeta'})\right] = f_{\zeta'} (1 - f_\zeta)$, Eq. (17)
takes the form of the well-known Kubo-Greenwood formula.

Apart from their use in Ref. 3 for the 1D modulation case, the  above formulas have also 
been succesfully applied to various
situations of electronic transport, 
such as hopping conduction \cite{11}a, Aharonov-Bohm effect \cite{11}b,
quantum Hall effect \cite{11}c, 
etc.

The resistivity tensor $\rho_{\mu\nu}$ is given in terms of the
conductivity tensor $\rho = \sigma^{-1}$. We will use the standard
expressions $\rho_{xx} = \sigma_{yy}/S, \rho_{yy} = \sigma_{xx}/S$,
and $\rho_{yx} = -\rho_{xy} = -\sigma_{yx}/S$ with
$S = \sigma_{xx}\sigma_{yy} - \sigma_{xy}\sigma_{yx}$.
 
\subsection{ Analytical evaluations}

The scattering mechanism enters the conductivity expressions (15) and
(16) through the relaxation time $\tau(E_\zeta)$ and the transition
rate $W_{\zeta\zeta'}$, respectively; in contrast, Eq. (17) is
independent of the scattering when the latter is weak \cite{11}.

We assume that the electrons are scattered elastically by randomly
distributed impurities. This type of scattering is dominant at the low
temperatures of the reported experiments. Further, we expand
the impurity potential in Fourier components, i.e., $U({\bf r} - {\bf R})=
\sum_{ {\bf q}} U_{\bf q} \exp\left[ i{\bf q}.({\bf r} - {\bf R})\right ]$,
with $U_{\bf q} = 2\pi e^2/\epsilon (q^2 + k^2_s)^{1/2}$ corresponding to the
screened impurity potential $U({\bf r}) = (e^2/\epsilon r)\exp(-k_sr)$;
${\bf r}$ and ${\bf R}$ are the electron and impurity positions,
respectively, ${\bf q}=q_x\hat{x}+ q_y\hat{y}$,  $\epsilon $ is the dielectric constant,
and $k_s$ the screening wave vector.

{\it Diffusive contribution}.
For weak modulation potentials $V_x$ and $V_y$, which is
pertinent to most of the reported experiments, we may use
$\mid \zeta> = \mid n,k_y,\xi>$ to evaluate the velocity
matrix elements appearing in Eq. (15); the latter are given by
Eqs. (5) and (6) for {\it rectangular} symmetry and by
Eqs. (11) and (12) for {\it hexagonal} symmetry. As for the
relaxation time $\tau(E_\zeta )$,  it is  defined by $1/\tau(E_\zeta) =
\sum_{\zeta'} W_{\zeta\zeta'}(v_\zeta-v_\zeta')/v_\zeta$.
Though the Landau levels broaden into bands, this definition fails at
the flat-band conditions, when $v_\zeta=v_\zeta'=0$. For this reason
we estimate it from the lifetime  given by $1/\tau(E_\zeta) =
\sum_{\zeta'} W_{\zeta\zeta'}$. In the limit $k_s \gg q$, we
obtain $\tau = \tau(E_\zeta) \approx (\pi \ell^2\hbar^2/N_IU^2_0)^{1/2}$
where $N_I$ is the 2D impurity density and $U_0 \approx 2\pi e^2/\epsilon k_s$.
However, at weak magnetic fields we may use  $\tau$ as
constant and estimate it from the zero-field mobility $\mu$:
$\tau = \tau_0 = \mu m^*/e$.

We now use Eqs. (5), (6), and (15) with $\sum_k \rightarrow
(L_y/\pi) \int^{a_x/2\ell^2}_0 dk_y$ and $\sum_\xi \rightarrow (L_x/
\pi a_x) \int^{2\pi}_0 d\xi$. The result for $\sigma^{dif}_{xx}$ is
\begin{equation}
\sigma^{dif}_{xx} \approx  {e^2 \over h} {\beta\tau \over \hbar\pi a_x}
 K_y^2\ell^4 V_y^2 
\sum_n e^{-u_y}\left[ L_n(u_y)\right]^2 \int_0^{2\pi}d\xi\int_0^{a_x/\ell^2}dk_y
f_{nk_y\xi}(1- f_{nk_y\xi})\sin^2\xi.
\end{equation}
The component $\sigma_{yy}^{dif}$ is given by Eq. (18) with $x$ and $y$
interchanged, and $\sin^2\xi$ replaced by $\sin^2(\ell^2K_xk_y)=\sin^2(K_xx_0)$.
In the limit of vanishing $V_y$ Eq. (18) gives the result of a 1D
modulation, $\sigma^{dif}_{xx} = 0$.
If we neglect the weak $k_y$- and $\xi$-dependence of the
factor $f_{nk_y\xi}(1 - f_{nk_y\xi})$ we obtain  the simplified expression
\begin{equation}
\sigma^{dif}_{xx} \approx   {e^2 \over h} {\beta\tau \over \hbar}
K_y^2  \ell^2 V_y^2 
\sum_n e^{-u_y} \left[ L_n(u_y)\right]^2 f_{n}(1 - f_{n}) . 
\end{equation} 
The corresponding  expression for  $\sigma_{yy}^{dif}$ is given by Eq. (19)
with $x$ and $y$ interchanged. 

For {\it hexagonal} symmetry we use 
$K_x = 2\pi/a$ and $K_y = 2\pi/\sqrt{3}a$. The results for $\sigma^{dif}_{xx}$
and $\sigma^{dif}_{yy}$
are similar to Eq. (18) and can be easily obtained using Eqs. (11) and (12 )
for the velocities. The result for $\sigma^{dif}_{yy}$, corresponding to
Eq. (19), is given by Eq. (19) with $K_y^2V_y^2$ replaced by 
$K_x^2V_x^2/2$
and that for $\sigma^{dif}_{xx}$ by 

\begin{equation}
\sigma^{dif}_{xx} \approx\pi {e^2 \over h} {\beta\tau \over \hbar}
K_y^2V_y^2
\left( 1 +{V^2_x \over 2V^2_y}   \right) e^{-u} \sum_n \left [ L_n(u)\right ]^2
f_{n}(1 - f_{n}), 
\end{equation}
where $u = 8\pi^2\ell^2/3a^2$. 

{\it Collisional contribution}.
To evaluate this contribution to order $V^2_\mu$ we must
use the perturbed wave function to order $V_\mu$. The procedure
for evaluating Eq. (16) is identical with that corresponding
to the 1D modulation detailed previously \cite{3}. We have again
$<\zeta\mid x \mid \zeta> - <\zeta' \mid x \mid \zeta'> = \ell^2
(k_y - k_y')$; the only new ingredient are the following matrix
elements
\begin{equation}
<nk_y\xi \mid y \mid nk_y\xi> = - \xi/K_y 
\end{equation}
and

\begin{equation}
\mid < nk_y\xi \mid e^{i{\bf q}\cdot{\bf r}}\mid n'k_y'\xi'>\mid^2
= (n!/n'!) u^{n'-n} e^{-u}
\left[ L^{n'-n}_n (u)\right] ^2
\delta_{\xi,\xi'+ c_yq_x } \delta_{k_y,k'_y-q_y},
\end{equation}
where $u = \ell^2(q^2_x + q^2_y)/2$ and $c_y=\ell^2K_y$.

We now use Eqs. (16), (20)-(21),  and the standard expression for the transition rate 
\begin{equation}
W_{\zeta\zeta'}=\sum_{{\bf q}} U_{{\bf q}}^2\mid < nk_y\xi \mid e^{i{\bf q}\cdot {\bf r}}\mid n'k_y'\xi'>\mid^2
\delta(E_{nk_y\xi}- E_{n'k'_y\xi'}).
\end{equation} 
We use the spectrum (3) and shift the argument of the cosines by $\ell^2K_xK_y=2\pi\alpha$, $\alpha$ integer, in the $\delta$ function as well as in the 
factor $f_{nk_y\xi}(1- f_{n'k'_y\xi'})$. Then  Eq. (16) takes the form 
\begin{eqnarray}
\sigma^{col}_{yy} &\approx& {e^2 \over h} {\beta N_IU^2_0 \over 4  a_x}
\nonumber
\sum_{n,n'} \int_0^{\infty}du\ e^{-u}u^{n'-n+1} [ L_{n}^{n'-n}(u)]^2\\  
\nonumber
\quad\quad&\times& \int_0^{2\pi}d\xi\int_0^{a_x/\ell^2}dk_y
\ f_{nk_y\xi}(1- f_{n', k_y+K_y+q_y,\xi-c_y(K_x+q_x)})\\*
\quad\quad&\times&\delta(E_{nk_y\xi}- E_{n', k_y+K_y+q_y,\xi-c_y(K_x+q_x)}).
\end{eqnarray}
We  proceed as follows. For weak magnetic fields involved in the problem the Landau-level index $n$ is large and the major contributions to the   sum
over $n'$ come from $n'$ values close $n$. With the asymptotic expansion of the Laguerre
polynomials, $e^{-u/2}  L_{n}(u)\approx (\pi^2nu)^{-1/4}\cos(2\sqrt{nu}-\pi/4)$, it's an excellent approximation to take $F_n(u_\mu)\approx F_{n'}(u_\mu)$. Then the
$\delta$ function becomes
\begin{eqnarray}
\nonumber
\delta(E_{nk_y\xi}&-& E_{n',k_y+K_y+q_y,\xi-c_y(K_x+q_x)})\approx \delta[(n-n')\hbar\omega_c\\*
\nonumber
&+&2F_n(u_x)V_x\sin c_x(K_y-q_y/2)\sin c_x( k_y+K_y-q_y/2)\\*
&+&2F_n(u_y)V_y\sin c_y(K_x-q_x/2)\sin c_y( k_x+K_x-q_x/2)].
\end{eqnarray}
The shift  by $\ell^2K_xK_y=2\pi\alpha$, $\alpha$ integer, in Eq. (25)  and in the factor $f_{nk_y\xi}(1- f_{n'k'_y\xi'})$   was made to stress the  formal  validity of Eqs. (24) and (25) for  $\alpha$ integer. If we don't make it, we must 
put $K_x=K_y= 0$ in the sine factors and change $E_{n',k_y+K_y+q_y,\xi-c_y(K_x+q_x)}$ to $E_{n',k_y+q_y,\xi-c_yq_x}$ wherever it appears. For 
$\alpha$ close to an integer though, one can reinstate $K_x$ and $K_y$ in Eqs. (24) and (25) as shown.

We now remark that the largest contribution to the integral over $u$ in Eq. (24) comes from very small
values of $q_x$ and $ q_y$  due to the factor $\exp(-u)$  or the factor $ 1/\sqrt{\pi^2nu}$  in the asymptotic expression $e^{-u} [ L_{n}(u)]^2\approx  
\cos^2(2\sqrt{nu}-\pi/4)/\sqrt{\pi^2nu}$. 
In addition, for the usual 2D systems
we have $k_s\approx 10^8$/m which is much larger than these small values of $q_x$ and $ q_y$ .  
With that in mind and in order to reduce the numerical work, 
we replace the $\delta$ function (25) by a Lorentzian of width $\Gamma$ and  neglect  in it and in the factor $f_{nk_y\xi}(1- f_{n', k_y+K_y+q_y,\xi-c_y(K_x+q_x)})$ the terms $\propto q_x$ or $\propto q_y$. 
Alternatively, we may expand the $\delta$ unction in powers of $q_x$ and $q_y$; then by far the leading contribution   comes from the zero-order
term given by Eq. (25) with $q_x=q_y=0$.
In addition, we neglect the term $q^2$ in $U_{\bf q}$. Further, from the sum over $n'$ we consider only the terms $n'=n$ and $n'=n\pm 1$; the term $n'=n$ gives the dominant contribution, about 90\%. Then the integral over $u$
can be evaluated  and Eq. (24)  takes the form
\begin{eqnarray}
\nonumber
\sigma^{col}_{yy} &\approx& {e^2 \over h} {\beta N_IU^2_0 \Gamma \over 2\pi^2 a_x}
\sum_{n} \{(2n + 1) \int_0^{2\pi}d\xi\int_0^{a_x/\ell^2}dk_y\\*
&&\times [D_{n,n}+(n+1) D_{n,n+1} +n D_{n,n-1}]\}
\end{eqnarray}
where
\begin{equation}
D_{n,n'}=f_{nk_y\xi}(1- f_{n', k_y+K_y,\xi-c_yK_x})/[(E_{nk_y\xi}- E_{n', k_y+K_y,\xi-c_yK_x})^2+\Gamma^2]
\end{equation}
As a result, when $2\pi\ell^2 /a_x a_y=\Phi_0/\Phi$ is an integer, the second and third terms in the argument of the $\delta$ function in Eq. (25) vanish  and entail $n=n'$, i. e., {\it the response is strongest when one flux quantum passes through an integral number of cells} as observed \cite{7, 13}. In this case the factor $[(...)^2+\Gamma^2]$  in 
Eq. (27) becomes $\Gamma^2$.


A qualitative understanding of the enhancement of the collisional conductivity for integer
$\alpha=\Phi_0/\Phi$ is as follows. In this case only  scattering between the states
$|k_y, \xi\rangle$ and $| k_y+K_y+q_y, \xi+cy(K_x,q_x)\rangle$ is allowed,
cf. Eq. (24). These states correspond to cyclotron orbits  
separated by a distance $\alpha a_x$ which is a multiple of the lattice period. One example is shown in Fig. 5 for  two orbits that encircle a unit cell. As shown, the orbits
are in the same relative position with respect to the
modulation lattice and correspond to electron states of the same energy.
Since impurity scattering  is an elastic process that   
 leads to hopping between states of the same energy,
the hopping between such cyclotron orbits for integer $\alpha$ contributes the most
to the conduction and  enhances the collisional conductivity.  
On the other hand, for $\alpha$ not  an integer the  position  of the two orbits
involved in the scattering process relative to  the modulation lattice changes; accordingly the 
enhancement mentioned above  is weakened. 

\begin{figure}[tpb]
\vspace{-1.5cm}
\includegraphics*[width=90mm,height=120mm]{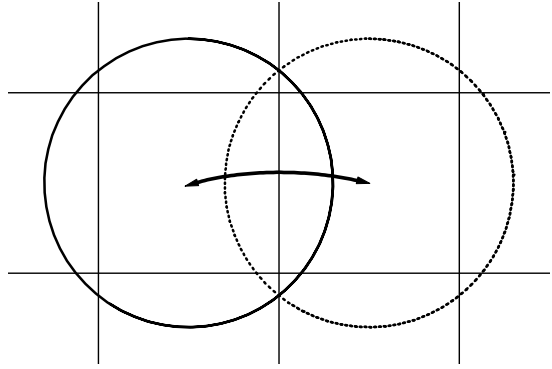}
\vspace{-2.5cm}
\caption{Scattering between two cyclotron orbits which encircle one unit cell.} 
\label{fig16}
\end{figure}

For those values of  the magnetic field for which $F_n(u_x)$ vanishes we use the
same wave functions and the spectrum   (4). If the modulation periods are the same,
we have $F_n(u_x)=F_n(u_y)=0$, $n\to n'$,  and Eq. (26) holds with $D_{n,n\pm 1}\to 0$.
If the modulation periods are not equal or if  $\Phi_0/\Phi$ is not an integer,
Eqs. (25) and (26) hold only approximately.
With all that in mind, the assumption that the small gaps are closed due to disorder,
and for computational convenience, we use Eq. (26) as an  approximation for all fields.

For the {\it hexagonal} modulation we obtain again Eq. (26) 
but now the energy spectrum is given by Eq. (10). Further,
$ a_x$ and $ u_y$ are replaced by
$ a$ and $u= 8  n^2 \ell^2/3a^2$,  
respectively. For $\sigma^{col}_{xx}$ the result is given by  Eq. (26) 
with $a_x$  replaced by $ a$; $u$  remains the same.

{\it The Hall conductivity}.
The evaluation of Eq. (17) for $\omega=0$ is readily performed with the
states $\mid nk_y\xi>$ and the energy levels given by Eqs. (3) or (10).
The only difference with the previous \cite{3} calculation is that a
factor $\exp[i(k_y-k_y^{'})/K_y]\delta_{\xi,\xi'}$ appears on the rhs of
Eq. (17) of Ref. 3 now written as $<nk_y\xi \mid V_\mu \mid n^{'}k_y^{'}\xi'>$.
For {\it rectangular}
symmetry we obtain $(\sigma_{yx}(0) \equiv \sigma_{yx})$ 
\begin{equation}
\sigma_{yx} = {e^2 \over h} {2\ell^2 \over \pi a_x} \sum_n (2n+1)
\int^{a_x/2\ell^2}_0 dk_y \int^{2\pi}_0 d\xi
{f_{nk_y\xi} - f_{n+1,k_y\xi} \over [1 + \lambda_{nx} \cos(K_xx_0)+
\lambda_{ny} \cos \xi]^2}, 
\end{equation}
where $\lambda_{n\mu} = V_\mu e^{-u_\mu/2} L^{-1}_{n+1}
(u_\mu)/\hbar \omega_c, \mu = x,y$. We notice that for $V_y = 0$ we
obtain the previous 1D result \cite{3}. We also remark that Eq. (28) is valid for
hexagonal symmetry with $a_x \rightarrow a$, $u_x = u_y = u =
8\pi^2\ell^2/3a^2$ and of course the different energy levels (Eq. (10))
that enter the factor $f_{nk_y\xi}$.

\section{Numerical results}

We now present  results for the   various resistivity components $\rho_{\mu\nu}$ using the
standard expressions given at the end of  Sec. III. A, and evaluating numerically the
conductivities given by Eqs. (18), (26),  and (27). For Figs. 3-7, and 10 we use the 
parameters of Ref. 8. They are: electron density $n_s=4.5\times 10^{15}/m^2$, temperature 
$T=5.5$ K or $T=1.6$ K, $a_x=a_y=804$\AA, and mobility $\mu=70\ m^2$/Vs. The corresponding 
parameters for Figs. 8-9, taken from Ref. 5, are $n_s=5.1\times 10^{15}/m^2$, temperature 
$T=4.2$ K, $a_x=a_y=2820$\AA, and mobility $\mu=140\ m^2$/Vs.
The relaxation time at zero magnetic field $\tau_0$ is estimated from the sample mobility
as $\tau_0=m^*\mu/e$. Then the level width is 
$\Gamma=(eBN_IU^2_0/\pi\hbar)^{1/2}=(e\hbar/m^*)\sqrt{B/\pi\mu}$.

In Figs. 3   
we plot  $\rho_{xx}$  and $\rho_{yy}$ as function of the magnetic field $B$ with $V_x=0.5$ 
meV for constant $\tau=\tau_0$.
As indicated, the various curves correspond to different $V_y$ and the dotted one, marked 1D, represents the 1D limit obtained with $V_y=0$.
The prominent peaks in the 2D case, marked by the integral value of $\alpha$, result
from the collisional contribution to the conductivity. The smaller peaks, between these
values of $\alpha$, correspond to the commensurability or Weiss oscillations.
Notice how the prominent peaks  of $\rho_{xx}$, in the 2D case,
remain rather insensitive to changes in $V_y$: this is so because they result
from the collisional contribution
$\sigma_{yy}^{col}$ which depends very weakly on $V_y$ through the energy spectrum.
The apparently drastic difference between the two figures results from the fact that
$\sigma_{\mu\mu}\ll \sigma_{xy}$ makes $S$ change little and
$\rho_{xx} = \sigma_{yy}/S$ while $\rho_{yy} = \sigma_{xx}/S$.
Upon reducing $V_y$ the contribution $ \sigma_{xx}^{dif}\sim V_y^2$,
given by Eq. (18), is affected drastically whereas $ \sigma_{yy}^{col}$ is not.

\begin{figure}[tpb]
\vspace{-1.5cm}
\includegraphics*[width=100mm,height=110mm]{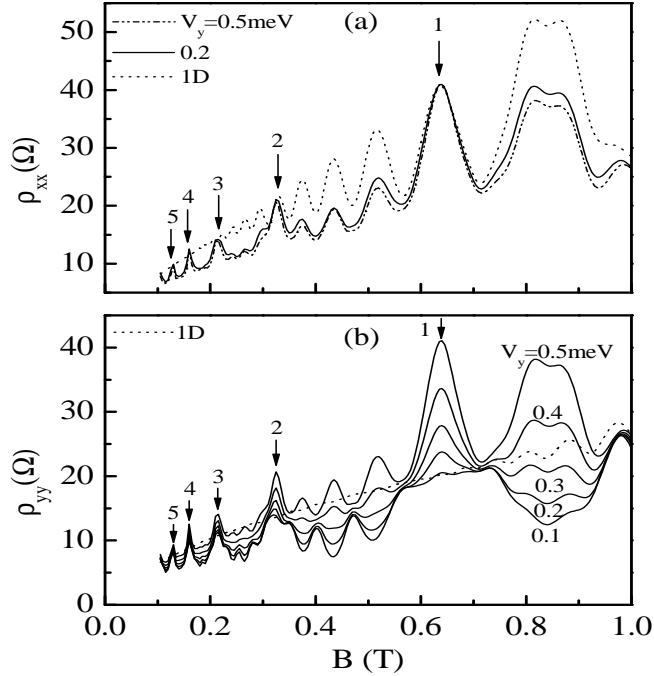}
\caption{Resistivity components $\rho_{xx}$  and $\rho_{yy}$ as a function of the
magnetic field $B$ for fixed $V_x=0.5$ meV  and varying $V_y$.
The dotted curve is the 1D limit  obtained with $V_y=0$.
The prominent peaks in the 2D case are marked by the integral value of $\alpha=\Phi_0/\Phi$.}
\label{fig3}
\end{figure}

In Figs. 4 and 5 we plot  again $\rho_{xx}$  and $\rho_{yy}$ as function of the
field $B$ with $V_x=V_y=0.5$ meV,
for the same  $\tau$ and $\Gamma\propto B^{1/2}$ as in Fig. \ref{fig3}  but with the
period $a_y$ being doubled from panel to panel as indicated.
The solid curves give the total resistivity, the dashed  ones the diffusive contribution,
defined by $\rho_{\mu\mu}^{dif} = \sigma_{\nu\nu}^{dif}/S$,
and the dotted  ones the collisional contribution, defined by
$\rho_{\mu\mu}^{col} = \sigma_{\nu\nu}^{col}/S$.
Notice how the prominent peaks move to lower fields with increasing $a_y$
as explained after Eq. (26); in panel (d) they have disappeared.
The 1D limit shown in panel (d) is obtained with $V_y=0$
and the difference in the $B$ dependence between
$\rho_{xx}$ and $\rho_{yy}$ is related to that of the corresponding conductivity
contributions. One of them, $\sigma_{xx}^{dif}$ given Eq. (18),
is affected drastically by changing $V_y$ and/or the period $a_y$
which enters
the factor $\sin^2\xi$, the others very weakly. 

\begin{figure}[tpb]
\vspace{-3.5cm}
\includegraphics*[width=140mm,height=120mm]{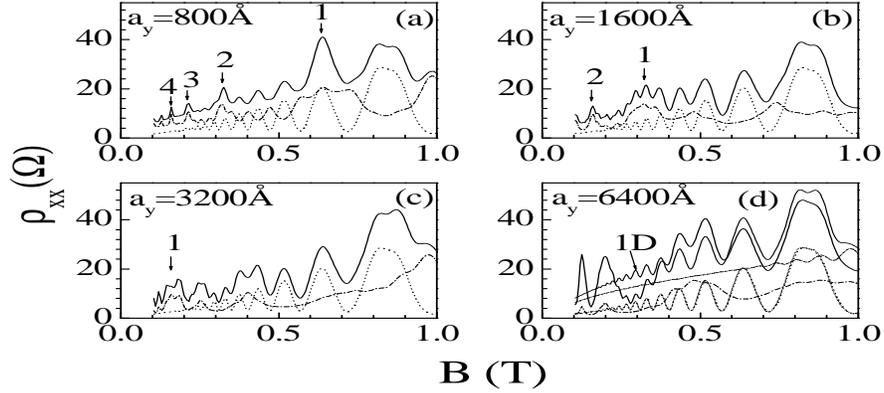}
\vspace{-3cm}
\caption{Resistivity component $\rho_{xx}$ as a function of the magnetic
field $B$ for fixed $V_x=V_y=0.5$ meV
and different periods $a_y$ as indicated. The solid curves give the total resistivity,
the dotted ones the diffussive contribution, and the dash-dotted ones the
collisional contribution.
The thinner curves in panel (d) are for the 1D limit ($V_y=0$.)
The prominent peaks in the 2D case are marked by the integral values of $\alpha$.}
 \label{fig4}
\end{figure}
\begin{figure}[tpb]
\vspace{-3.5cm}
\includegraphics*[width=140mm,height=120mm]{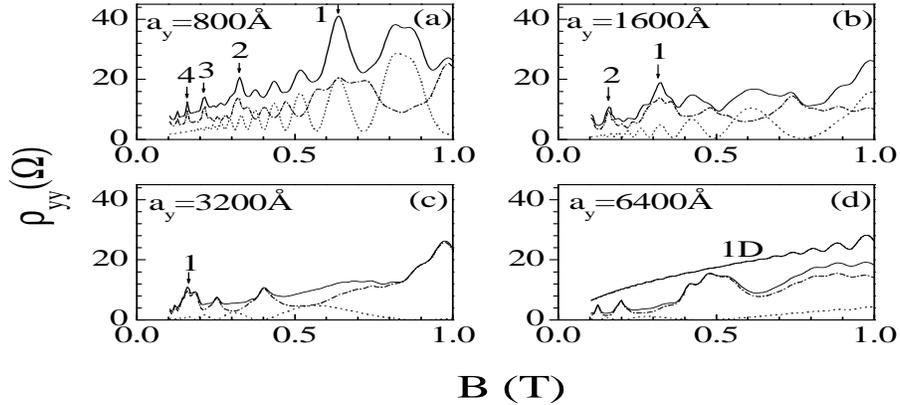}
\vspace{-2.5cm}
\caption{The same as in Fig. \ref{fig4} but for the resistivity component $\rho_{yy}$.} 
\label{fig5}
\end{figure}

\begin{figure}[tpb]
\vspace{-1.5cm}
\includegraphics*[width=100mm,height=110mm]{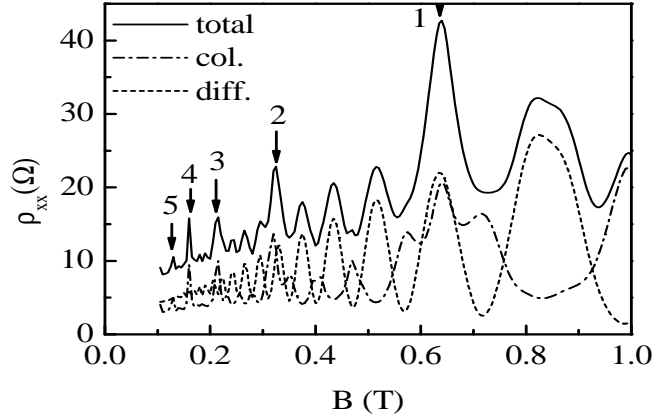}
\vspace{-3cm}
\caption{Resistivity component $\rho_{xx}$ as a function of the magnetic field $B$
for fixed $V_x=V_y=1$ meV and $T=5.5$ K.
The solid curves give the total resistivity, the dotted  ones the
diffussive contribution, and the dash-dotted
ones the collisional contribution. The prominent peaks are marked by the
integral values of $\alpha
=\Phi_0/\Phi$.}
\label{fig6}
\end{figure}

We now look more closely at the experimental results of Refs. 8 and 5 with
the parameter sets specified above.
The parameters $V_x$ and $V_y$ are not known. In Fig. \ref{fig3} we have shown the total
resistivity for $V_x=V_y=0.5$ meV, constant $\tau=\tau_0$. In Fig. \ref{fig6} we plot again
$\rho_{xx}$ versus $B$ but now
we show the contributions $\rho_{xx}^{dif}$ and  $\rho_{xx}^{col}$ as well.
In addition, we take $V_x=V_y=1$ meV, and $1/\tau\sim\Gamma\sim B^{1/2}$.
Because  $\sigma_{\nu\nu}^{dif}\ll \sigma_{\nu\nu}^{col}$, the difference in
the total resistivity is very small between the two sets of  modulation strengths.
However, the oscillation amplitudes
in $\rho_{\mu\mu}^{col}$ are higher in the present case and $\rho_{xx}$ increases
more slowly with $B$ as observed
 \cite{7}. Upon closer inspection we see that the prominent peaks,
marked by the integral values of $\alpha$, result entirely from the collisional
contribution $\sigma_{\nu\nu}^{col}$. As can be seen in Fig. \ref{fig2} of the next article
\cite{13}, the amplitudes and positions of these peaks agree well to very well with the
experimental results. Notice also how  the Weiss oscillations of $\rho_{xx}^{dif}$ and
$\rho_{xx}^{col}$, between these peaks, are in antiphase. These experimental results for
$\rho_{yy}$ and   
$\rho_{xx}$  are for two orthogonal crystal directions $[011]$ and
$[01\bar1]$ taken from different samples. They have similar structures and the curves
 may be fitted theoretically using slightly different potentials. A direct comparison between
experimental and theoretical results is made in Fig. \ref{fig6} of the next article \cite{13} for 
$\tau=\tau_0$ but qualitatively the agreement is the same for $\tau\propto B^{-1/2}$.
Notice, however, that the theoretical oscillation amplitudes and the overall value of $\rho_{xx}$ in the low-$B$ region, below  $\alpha=2$, agree less well
with the experimental ones than those in the high-$B$ region.

\begin{figure}[tpb]
\vspace{-2.8cm}
\includegraphics*[width=110mm,height=150mm]{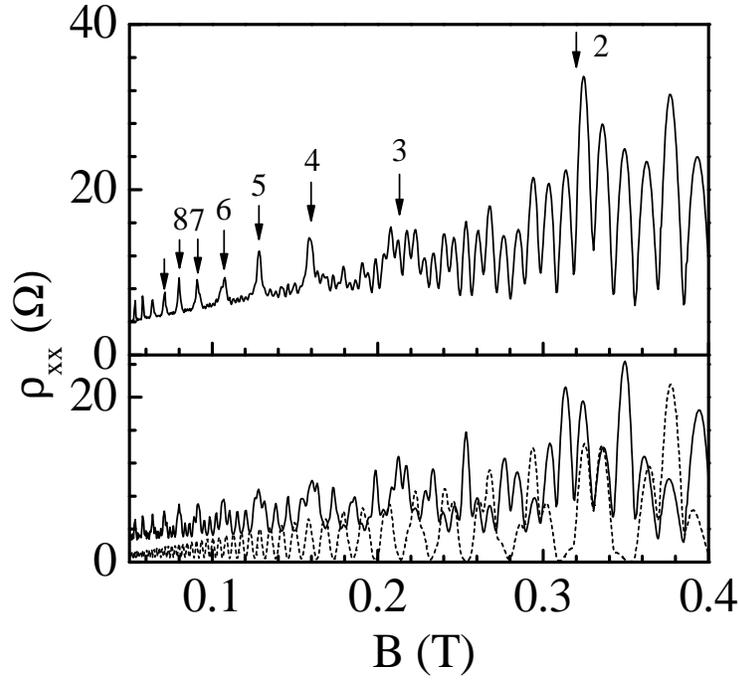}
\vspace{-3.5cm}
\caption{The same as in Fig. \ref{fig6} but for temperature $T=1.6$. The lower panel shows
the collisional contribution
(solid curve) and the diffusive one (dotted curve); the upper panel shows their sum. } 
\label{fig7}
\end{figure}

\begin{figure}[tpb]
\vspace{-2.8cm}
\includegraphics*[width=110mm,height=150mm]{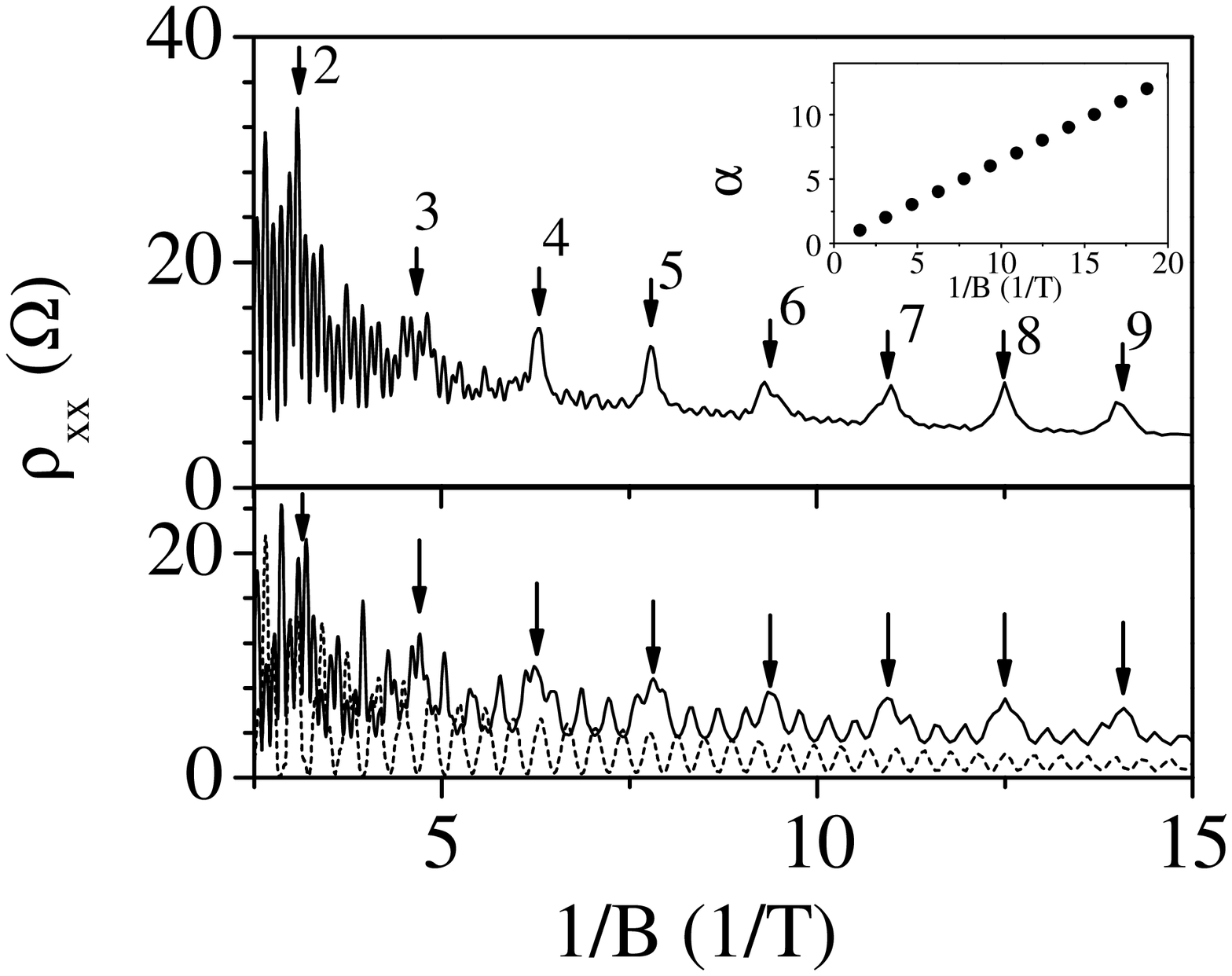}
\vspace{-4.5cm}
\caption{Resistivity component $\rho_{xx}$ as a function of inverse magnetic field $1/B$.
The curves are marked as in
Fig. \ref{fig7}. The inset shows the peak position versus $1/B$. } 
\label{fig8}
\end{figure}

In Ref. \cite{13} results are given for temperature 1.6 K and otherwise the same parameters.
We show the calculated $\rho_{xx}$ for this case in Fig. \ref{fig7}. As can be seen,
lowering the temperature makes visible all  prominent peaks marked by arrows
for $\alpha=1,...,8$.
Their positions occur at fields $B=0.64, 0.32, 0.21, 0.16, 0.13, 0.11, 0.09, 0.08$ T
and compare very well with the experimental ones,  see Fig. \ref{fig3} in the next article
\cite{13}. To see more clearly the oscillations we replot,
in Fig. \ref{fig8}, the resistivities in the low-field region  of Fig. \ref{fig7} as a function of $1/B$. 

\begin{figure}[tpb]
\vspace{-1.8cm}
\includegraphics*[width=110mm,height=150mm]{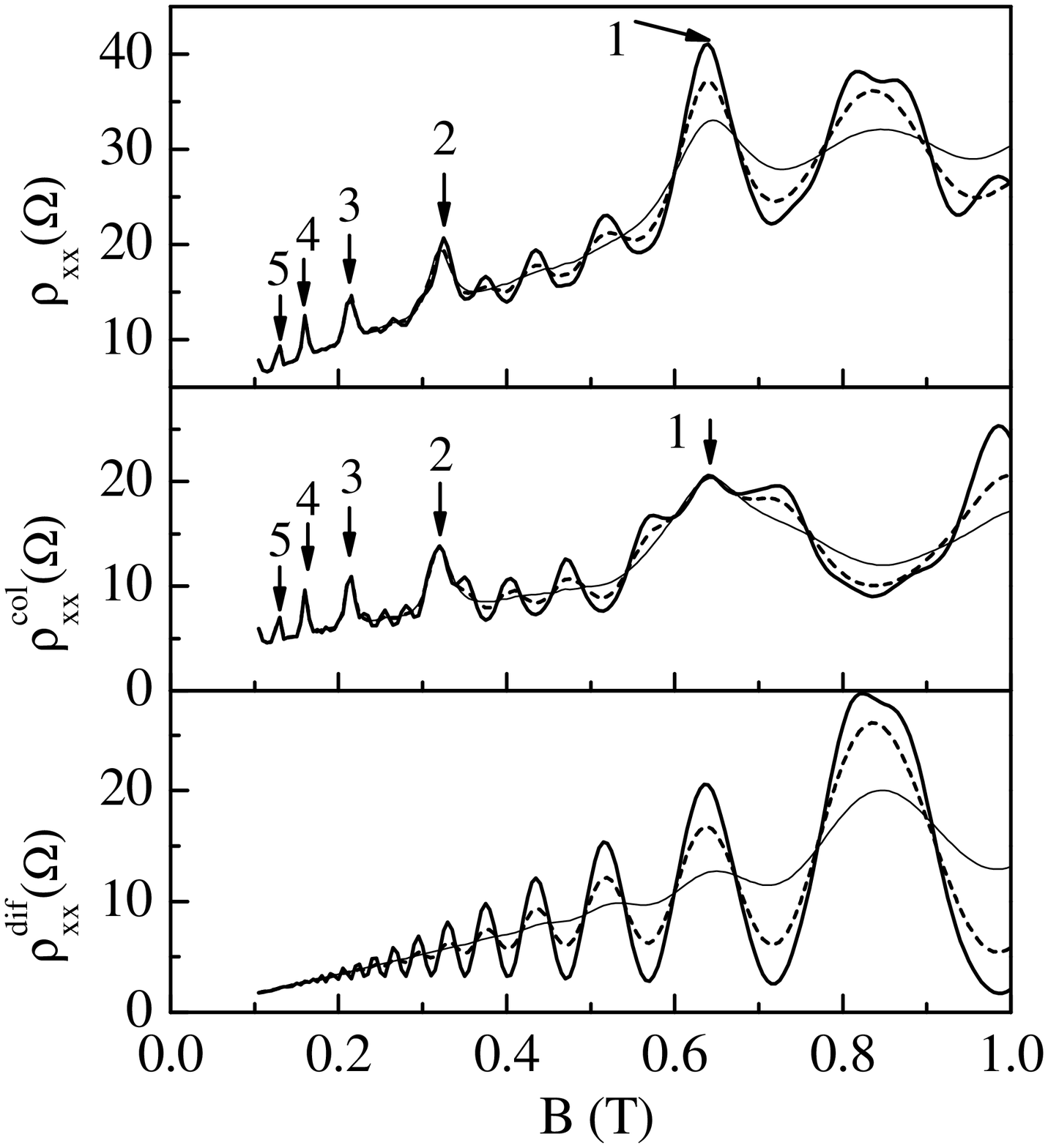}
\vspace{-3.5cm}
\caption{Resistivity component $\rho_{xx}$ as a function of the magnetic field $B$.
The solid, dotted, and thin solid curves correspond to $T=5, 10, 20$ K, respectively.}
\label{fig9}
\end{figure}

The temperature dependence of the oscillations  is shown in Fig. \ref{fig9}. The solid, dotted,
and thin solid curves correspond to $T=5, 10, 20$ K, respectively. As can be seen,
these new oscillations are more robust than the Weiss oscillations and persist
at $T=20$ K. However,
their damping with $T$ is weaker than the observed  one \cite{13}.

\begin{figure}[tpb]
\vspace{-1.5cm}
\includegraphics*[width=100mm,height=110mm]{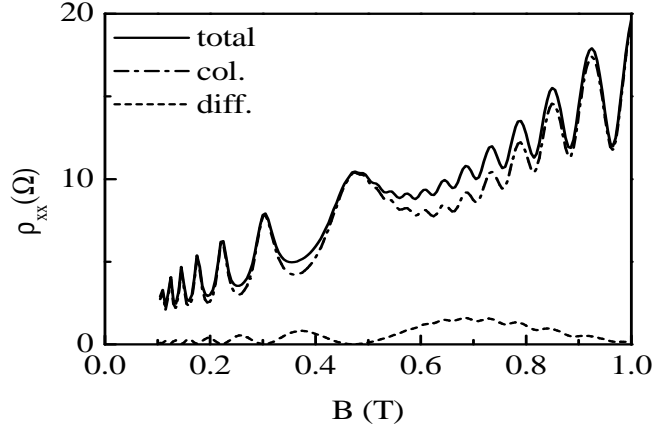}
\vspace{-2.5cm}
\caption{The same as in Fig. \ref{fig6} with the parameters of Ref. 5 and $V_x=V_y=0.2$ meV.}
\label{fig10}
\end{figure}

In Fig. \ref{fig10} we  plot   $\rho_{xx}$ in the manner of Fig. \ref{fig6} but for the parameters of
Ref. 5 involving the much longer
periods $a_x=a_y=2820$\ \AA. The modulation strengths are  $V_x=V_y=0.2$ meV and very
close to those used in Refs. 5 and 6. The agreement with the experimental 2D results of
Ref. 5 is very good: below approximately $B=0.5$ T we have the Weiss oscillations and
above it the Shubnikov-de Haas ones.  One noticeable feature here is the absence of
the prominent peaks for integral values of $\alpha$. This is so because the much longer
periods involved make  $\alpha=2\pi\ell^2/a_xa_y$ integer for much smaller  values of $B$.
For instance, $\alpha=1$ occurs
at $B=0.05$ T and  the corresponding peak is not resolved. The agreement is also as good if
we use $a_x=a_y=3650$\ \AA \ and otherwise the same parameters pertaining to another sample.

\begin{figure}[tpb]
\vspace{-1.5cm}
\includegraphics*[width=100mm,height=110mm]{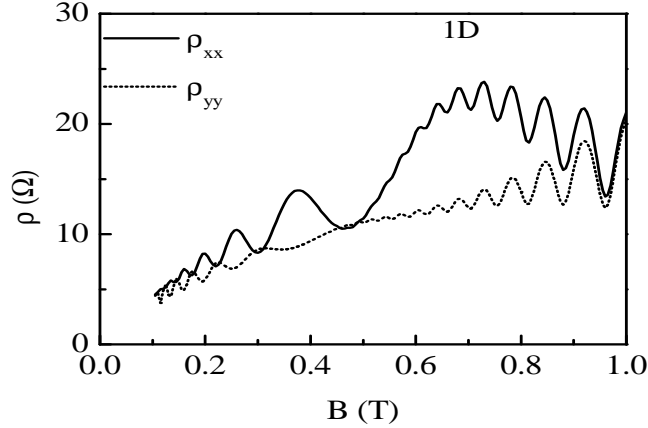}
\vspace{-2.5cm}
\caption{The 1D resistivity component $\rho_{\mu\mu}$ as a function of the
magnetic field $B$ obtained with
$a_x=2820$\ \AA, $V_x=0.5$ meV, and $V_y=0$.} 
\label{fig11}
\end{figure}

Reference 5 reported  results also for 1D modulations. As mentioned earlier, we can
obtain the 1D limit from the present 2D results by considering   a vanishing $V_y$.
In Fig. \ref{fig11} we show the 1D limit of the total $\rho_{xx}$ and $\rho_{yy}$ for
$a_x=2820$\ \AA, $V_x=0.5$ meV, and $V_y=0$.
Although the agreement between theory and experiment is very good,
it must be noticed that it is obtained with $V_x=0.5$ meV and not $V_x=0.2$ meV
that we used in Fig. \ref{fig10}. Since the the 1D or 2D modulations are produced
by illumination of the samples, we expect them to have the same strength.
If we use $V_x=0.2$ meV we can obtain good agreement if we use a $\tau$ smaller
by about a factor of 2  in Eq. (18). As stated in Refs. 5 and 6, this may be an
indication that in this very high mobility samples the fine structure of the
energy spectrum, that the present theory
neglects, is partially resolved. However, the experimental data was taken at
$T=4.2$ K and, as no such fine structure has been observed above mK temperatures,
alternative explanations have been proposed \cite{4}, \cite{13}.

\begin{figure}[tpb]
\vspace{-1.5cm}
\includegraphics*[width=100mm,height=110mm]{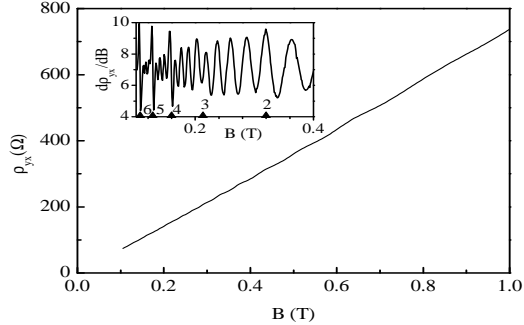}
\vspace{-5cm}
\caption{The Hall resistivity $\rho_{yx}$ versus magnetic field $B$
with the parameters of Fig. \ref{fig6}. The inset shows the derivative $d\rho_{yx}/dB$  versus $B$.} 
\label{fig12}
\end{figure}

Finally, in Fig. \ref{fig12} we show the Hall  resistivity $\rho_{yx}$ for the parameters
of Fig. \ref{fig6}. As in the case of 1D modulations, it exhibits very weak oscillations.
They are better seen in the inset which shows the derivative $d\rho_{yx}/dB$  versus $B$.
The triangles on the $x$ axis mark the positions of the integral values of $\alpha$ for
which enhanced oscillations are observed.

\section{Concluding remarks}

    We presented a theory of magnetotransport in 2D
superlattices using the energy spectrum and wave functions that result from the tight-binding difference equation when the parameter $\alpha=\Phi_0/\Phi$ is an integer. As emphasized in the text and supported with the results for the DOS shown in Fig. 4,  the description holds approximately for all fields if we assume that the small  gaps in the energy spectrum are closed due to disorder. The reasonable-to-good agreement with the experimental results strongly supports this assumption. 

As detailed in the text, the prominent peaks, for $\alpha=\Phi_0/\Phi$ integer, result from the collisional contribution to the conductivity $\sigma_{yy}^{col}$, require sufficiently {\it short} periods, and depend very weakly
on the value of the modulation strengths   $V_{x}$ and  $V_{y}$. Upon increasing the period along one direction we showed how they move to lower fields. Accordingly, for periods between $3000$\ \AA \ and  $4000$\ \AA \  these peaks occur at much smaller magnetic fields and are not resolved \cite{5}. The agreement between our results and the experimental ones, as presented in Ref. 8 and detailed in the next article  \cite{13}, is  good for the peak positions at all fields.  The oscillation amplitudes agree well at  relatively
high fields but less well at low fields. As shown in Fig. \ref{fig9}, these oscillations are quite robust with respect to the temperature but their damping with temperature is weaker than
the observed one. 

Between the oscillations for $\alpha=\Phi_0/\Phi$ integer  we have the Weiss oscillations.
The relative phase between those of $\rho_{xx}$ and those of $\rho_{yy}$ depends on the
values
of the modulation strengths, cf. Fig. \ref{fig3} in which the period is the same for all curves,
and of the modulation periods, cf. Figs. 4 and 5 in which the modulation strengths are the
same for all panels. We notice that as $V_y$ becomes smaller and smaller than $ V_x$, the
oscillations resemble more
closely those corresponding to 1D
{\it weak} modulations \cite{14}, cf. Fig. \ref{fig3}. The results for the latter can be extracted
from the present 2D ones if we take the   modulation strength  along one direction to be
zero.  
    
The relative phase between $\rho_{xx}$ and  $\rho_{yy}$ for 1D and 2D modulations depends
strongly on the ratio of the modulation strengths   $V_{x}$ and  $V_{y}$.
Since one resistivity component vanishes for  1D modulations, this conclusion could not be
reached by studying only the latter. Similar results were reported in Ref. 6.

\leftline{\bf Acknowledgments}

This work was supported by the Canadian NSERC Grant No. OGP0121756,
the Belgian Interuniversity Attraction Poles (IUAP),
the Flemish Concerted Action (GOA) Programme, and the EU-CERION programme.
We also thank A. Long and J. H. Davies for stimulating discussions and important
clarifications concerning the experimental results of the next article.

\end{document}